\documentclass[amsmath,twocolumn,pra,longbibliography,superscriptaddress]{revtex4-2} 
\usepackage{amsthm,amssymb,amsfonts,graphicx,verbatim, xcolor,bm} % The basics
\usepackage{hyperref} % URL links
\usepackage[utf8]{inputenc} % encoder
\usepackage{siunitx} % nice units
\usepackage{ulem}

\newcommand{\Tr}{{\rm Tr}}

\begin{document}

\title{Keldysh field theory of dynamical exciton condensation transitions \\ in nonequilibrium electron-hole bilayers}
\author{Yongxin Zeng}
\affiliation{Department of Physics, Columbia University, New York, NY 10027}
\affiliation{Center for Computational Quantum Physics, Flatiron Institute, New York, NY 10010}
\affiliation{Department of Physics, University of Texas at Austin, Austin, TX 78712}
\author{Valentin Cr\'epel}
\affiliation{Center for Computational Quantum Physics, Flatiron Institute, New York, NY 10010}
\author{Andrew J. Millis}
\affiliation{Department of Physics, Columbia University, New York, NY 10027}
\affiliation{Center for Computational Quantum Physics, Flatiron Institute, New York, NY 10010}

\begin{abstract}
Recent experiments have realized steady-state electrical injection of interlayer excitons in electron-hole bilayers subject to a large bias voltage. In the ideal case in which interlayer tunneling is negligibly weak, the system is in quasi-equilibrium with a reduced effective band gap. Interlayer tunneling introduces a current and drives the system out of equilibrium. In this work we derive a nonequilibrium field theory description of interlayer excitons in biased electron-hole bilayers. In the large bias limit, we find that p-wave interlayer tunneling reduces the effective band gap and increases the effective temperature for intervalley excitons. We discuss possible experimental implications for InAs/GaSb quantum wells and transition metal dichalcogenide bilayers.
\end{abstract}

\maketitle

{\it Introduction.---} Excitons are bosonic bound states of conduction band electrons and valence band holes in semiconductors. The possibility of Bose-Einstein condensation of excitons was first proposed \cite{blatt1962bose,keldysh1968collective} over sixty years ago. It was later realized \cite{lozovik1976new} that condensation of interlayer excitons in bilayer two-dimensional systems has striking experimental consequences including counterflow superfluidity and Josephson-like tunneling peaks \cite{fogler2001josephson, stern2001theory}. Equilibrium interlayer exciton condensation has been experimentally established in quantum Hall bilayers \cite{eisenstein2004bose,eisenstein2014exciton,spielman2000resonantly,kellogg2004vanishing,tutuc2004counterflow}. Equilibrium exciton condensation in the absence of a magnetic field has been theoretically studied in a number of contexts \cite{zhu1995exciton,min2008room,perali2013high,fogler2014high,wu2015theory,zeng2022layer, conti2020doping, conti2021electron, conti2023chester}, but has so far remained elusive experimentally in conventional semiconductor systems despite much effort \cite{croxall2008anomalous,seamons2009coulomb,gorbachev2012strong,burg2018strongly,wang2019evidence}.

Group-VI transition metal dichalcogenides (TMDs) with chemical formula $MX_2$ (where $M={\rm Mo,W}$ and $X={\rm S,Se,Te}$) are a class of two-dimensional semiconductors that host strongly bound excitons \cite{mak2010atomically,he2014tightly,wang2018colloquium,mak2016photonics,regan2022emerging} and can be stacked in various combinations. When two TMD layers are stacked, electrons from one layer and holes from the other layer form interlayer excitons that are strongly bound even with thin insulating barriers separating the electron and hole layers. Interlayer excitons in TMD bilayers have long lifetimes and electrically tunable properties \cite{fang2014strong,rivera2015observation,jauregui2019electrical,mak2018opportunities}. If separate contacts are made on the electron and hole layers \cite{wang2019evidence,ma2021strongly,qi2023thermodynamic,nguyen2023perfect,qi2023perfect}, the chemical potentials of carriers in the two layers are controlled separately, and their difference, the bias voltage, controls the exciton chemical potential \cite{zeng2020electrically, xie2018electrical}. When the exciton chemical potential exceeds the lowest bound state energy of electron-hole pairs, interlayer excitons are electrically injected into the bilayer system and undergo Bose-Einstein condensation (BEC) at low enough temperatures. Excitonic insulating states in TMD bilayers have been established in recent experiments by compressibility measurements \cite{ma2021strongly, qi2023thermodynamic} and drag measurements \cite{nguyen2023perfect, qi2023perfect}.

If tunneling between layers is negligible, the potential difference required to maintain a nonzero steady state exciton density can be gauged away, so the system is equivalent to an equilibrium electron-hole bilayer with a reduced effective band gap. Nonzero interlayer tunneling introduces a tunneling current \cite{wang2019evidence,zeng2020electrically} that drives the system out of equilibrium, leading to new physics \cite{sun2023dynamical} different from that of driven-dissipative condensates \cite{sieberer2013dynamical, sieberer2016keldysh, dunnett2016keldysh, maghrebi2016nonequilibrium, dalla2013keldysh}. In this Letter we present a microscopic theory of the nonequilibrium exciton condensate based on a Keldysh nonequilibrium field theory \cite{keldysh1965diagram,kamenev2023field,altland2010condensed, sieberer2016keldysh} that includes the effects of both a bias voltage and interlayer tunneling.

%Depending on the symmetries of the bilayer crystal and the orbital character of each layer, interlayer tunneling may transform trivially (s-wave) or nontrivially (p-wave, d-wave, {\it etc.}) under rotations, or may be a spatially varying function. 
In TMD bilayers, because the conduction and valence band extrema are located at the $\pm K$-valleys, the functional form of interlayer tunneling depends on the local stacking registry \cite{wang2017interlayer,rivera2018interlayer,liu2015electronic,yu2017moire}. In this Letter we focus on the experimentally relevant case of angle-aligned TMD homobilayers in which interlayer tunneling is uniform in space, and assume in our explicit calculations p-wave interlayer tunneling that applies to most of the high-symmetry stacking registries of TMDs as well as InAs/GaSb quantum wells \cite{liu2008quantum, xue2018time, zeng2022plane}. Different from s-wave tunneling, p-wave interlayer tunneling produces a potential landscape that is second order in the phase angle of the exciton field, leading when no bias voltage is applied to a second-order Josephson effect \cite{sun2021second} that breaks the interlayer phase symmetry down to $\mathbb{Z}_2$ from $U(1)$. We find that in the large bias limit, the system is described by an effective action in which interlayer $U(1)$ phase symmetry is effectively restored, but p-wave interlayer tunneling leads to a reduced effective band gap and an increased effective temperature for intervalley excitons.

\begin{figure}
\centering
\includegraphics[width=\linewidth]{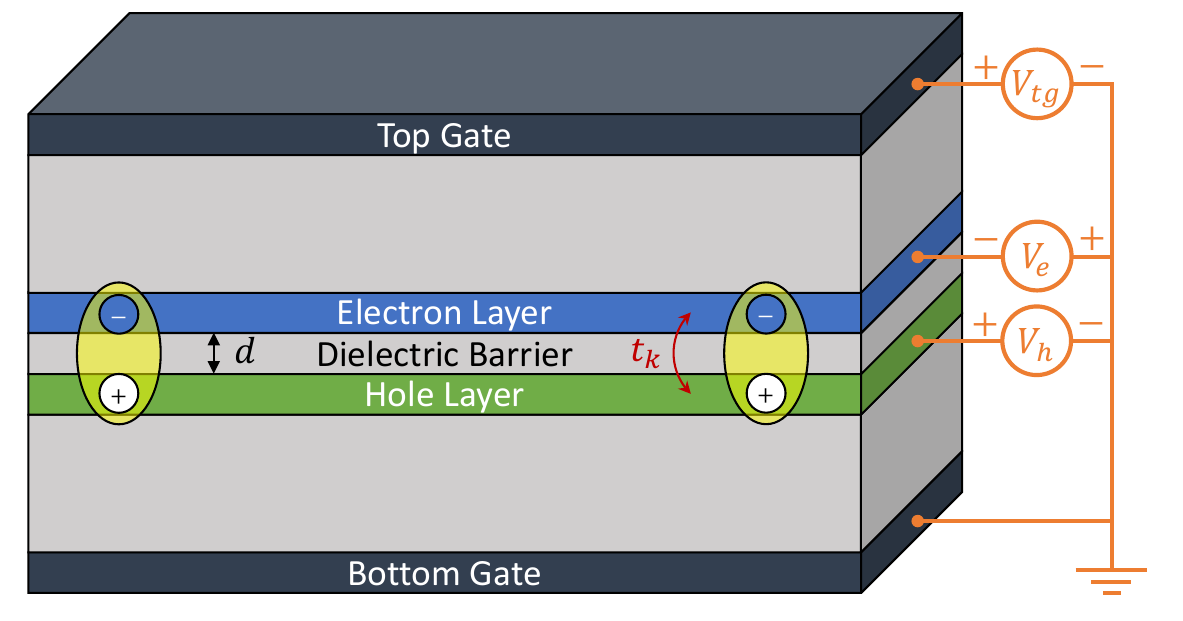}
\caption{An electrically controlled electron-hole bilayer. A negative voltage $-V_e$ and a positive voltage $+V_h$ are applied on the electron and hole layers respectively to inject electrons and holes into the system. The chemical potential of interlayer excitons $\mu_x=\mu_c-\mu_v = e(V_e+V_h)$ is set by the bias voltage between layers. The bottom gate is grounded, and the top gate voltage $V_{tg}$ produces a perpendicular electric field that tunes the band gap. The gray regions represent dielectric layers.}
\label{fig:device}
\end{figure}

{\it Model.---} We consider an electron layer and a hole layer separated by a weakly conducting barrier as shown in Fig.~\ref{fig:device}. Experimentally the system is controlled in two ways: by tuning the top/bottom gate potential difference (fundamentally an equilibrium effect) and by connecting the electron and hole layers to reservoirs held at different chemical potentials, enabling injection and removal of carriers. The system is described by the Hamiltonian $H = H_0 + H_t + H_C$
%\begin{equation} \label{eq:H}
%H = H_0 + H_t + H_C,
%\end{equation}
where 
\begin{equation} \label{eq:H_0}
H_0 = \sum_{\tau bk} \xi_{bk} a_{\tau bk}^{\dagger} a_{\tau bk}
\end{equation}
describes the kinetic energy of conduction band electrons and valence band holes. Here $\tau=\pm$ is the valley index and $b=c,v$ is the band (layer) index. $\xi_{ck} = k^2/2m_e^* + E_g/2$ and $\xi_{vk} = -k^2/2m_h^* - E_g/2$ describe the dispersion of the conduction and valence bands, where $m_e^*$ and $m_h^*$ are the effective masses of electrons and holes and $E_g$ is the band gap that can be tuned by a perpendicular electric field produced by the difference between top and bottom gate voltages. The next term in the Hamiltonian
\begin{equation} \label{eq:H_t}
H_t = \sum_{\tau k} t_{\tau k} a_{\tau ck}^{\dagger} a_{\tau vk} + {\rm h.c.}
\end{equation}
describes interlayer tunneling arising from hybridization of electron and hole wavefunctions in the two layers. A nonzero $t_{\tau k}$ explicitly breaks the $U(1)$ symmetry of the model associated with charge conservation in each layer. The momentum and valley dependence of $t_{\tau k}$ depends on symmetries of the system and is crucial for our upcoming results. For most of the high-symmetry stacking registries of angle-aligned TMD homobilayers, direct tunneling is forbidden by rotational symmetry \cite{rivera2018interlayer,liu2015electronic}, leading to p-wave interlayer tunneling
\begin{equation} \label{eq:t_k}
t_{\tau k} = v_t (\tau k_x + ik_y).
\end{equation}
This form of interlayer tunneling also applies to InAs/GaSb quantum wells \cite{liu2008quantum, xue2018time, zeng2022plane}, in which case $\tau$ is the spin index. In this Letter we focus on p-wave interlayer tunneling and briefly discuss other forms of interlayer tunneling at the end. The Coulomb interaction term
\begin{equation} \label{eq:H_int}
H_C = \frac{1}{2A} \sum_{bb'\tau\tau'} \sum_{kk'q} V_{bb'}(q) a_{\tau b,k+q}^{\dagger} a_{\tau' b',k'-q}^{\dagger} a_{\tau'b'k'} a_{\tau bk},
\end{equation}
where $A$ is the system area, distinguishes intralayer ($b=b'$) and interlayer ($b\ne b'$) interactions but neglects intervalley scattering due to the large momentum transfer required. Electron-hole exchange interactions \cite{yu2014dirac, glazov2014exciton, yu2014valley, wu2015exciton, qiu2015nonanalyticity} are also neglected due to the suppression of current matrix elements between electrons and holes in different layers.

By tuning the electrochemical potential of the electron layer $\mu_c = eV_e$ near the bottom of the conduction band and the hole layer $\mu_v = -eV_h$ near the top of the valence band, electrons and holes are injected into the system and form interlayer excitons. The chemical potential of interlayer excitons $\mu_x=\mu_c-\mu_v = e(V_e+V_h)$ is set by the bias voltage between two layers. %In the absence of interlayer tunneling, the electron and hole layers are independent and the potential difference $\mu_x$ can be eliminated using the $U(1)$ gauge symmetries corresponding to charge conservation in both layers separately. In this case the bias voltage $\mu_x$ effectively shifts the relative energy between two layers. In the presence of interlayer tunneling, sustaining the imposed chemical potential difference requires the creation of an electrical current between two layers. This drives the system out of equilibrium unless $\mu_x=0$. These non-equilibrium effects are the focus of our analysis.
 
{\it Keldysh action.---} %Starting from the Hamiltonian \eqref{eq:H}, 
We derive a nonequilibrium field theory that describes a biased electron-hole bilayer with interlayer tunneling based on the Keldysh formalism \cite{keldysh1965diagram,kamenev2023field,altland2010condensed, sieberer2016keldysh}, outlining here the procedure to obtain the Keldysh action and presenting the main results, with detailed derivations left for the Supplemental Material. We express the model as a path integral along a closed time path $C$ that starts from the distant past, proceeds to the distant future, and then returns to the starting point. The generating function is
\begin{equation} \label{eq:Z_def}
Z = \Tr\left\{ \rho_0 \mathcal{T}_C \exp[-i\int_C dt H(t)] \right\} \bigg / \Tr(\rho_0),
\end{equation}
where $\mathcal{T}_C$ is the contour ordering operator along $C$, and $\rho_0$ is the density matrix of the system in the distant past which we take as the equilibrium distribution of decoupled electron and hole layers: $\rho_0 = e^{-(H_0-\mu_c N_c - \mu_v N_v)/T}$ where $N_b = \sum_{\tau k} a_{\tau bk}^{\dagger} a_{\tau bk}$ is the number of electrons in each layer. For notational convenience Eq.~\eqref{eq:Z_def} is written for a closed system; coupling to leads is included in the theory as the imaginary (dissipative) part of inverse Green functions as detailed in the Supplemental Material. To derive a theory of excitons, we perform a Hubbard-Stratonovich transformation of interlayer electron-hole interactions and introduce the electron-hole pairing fields $\Delta_{kq}^{\tau\tau'}$, where $k$ and $q$ are respectively the relative momentum and center-of-mass momentum of an electron-hole pair, and $\tau, \tau'$ are the valley indices of electrons and holes. A nonzero value of $\Delta$ reflects spontaneous breaking of interlayer $U(1)$ symmetry associated with formation of the exciton condensate. The $k$-dependence of the pairing fields is irrelevant to the low-energy physics we discuss and is eliminated by projecting the $\Delta$ fields onto the 1s-exciton basis by defining
\begin{equation}
\Delta_{kq}^{\tau\tau'} = \frac 1A \sum_{k'} V_{cv}(k-k') \varphi_{k'} \Phi_q^{\tau\tau'},
\end{equation}
where $\varphi_k$ is the 1s-exciton wavefunction that is the lowest-energy solution of the eigenvalue equation
\begin{equation}
\frac{k^2}{2m}\varphi_k - \frac 1A \sum_{k'} V_{cv}(k-k') \varphi_{k'} = -E_b \varphi_k.
\end{equation}
Here $m=m_e^* m_h^*/(m_e^*+m_h^*)$ is the reduced mass of an electron-hole pair and the exciton binding energy $E_b$ is defined as the absolute value of the 1s-exciton energy. The 1s-exciton fields $\Phi$ have two valley indices, one for electrons and the other for holes, and we express them in terms of a four-component spinor $(\Phi^{\mu})$ defined as $\Phi^{\tau\tau'} = (\sum_{\mu} \Phi^{\mu}\tau_{\mu}/\sqrt{2})^{\tau\tau'}$ where $\tau_0$ and $\tau_{1,2,3}$ are the $2\times 2$ identity and Pauli matrices in valley space \cite{wu2015theory}. In this notation $\Phi^0,\Phi^3$ are intravalley exciton fields and $\Phi^1,\Phi^2$ are intervalley exciton fields. Integrating out the fermion fields, we obtain an effective action in terms of the 1s-exciton fields $\Phi$. Following the convention widely used in the literature on Keldysh field theory \cite{kamenev2023field, altland2010condensed, sieberer2016keldysh, keldysh1965diagram}, we transform the forward ($+$) and backward ($-$) branches of the $\Phi$ fields into the {\it classical} (c) and {\it quantum} (q) fields defined as
\begin{equation}
\begin{pmatrix} \Phi^{\rm c} \\ \Phi^{\rm q}
\end{pmatrix} = \frac{1}{\sqrt{2}} \begin{pmatrix} \Phi^+ + \Phi^- \\ \Phi^+ - \Phi^-
\end{pmatrix}.
\end{equation}
The generating function is now expressed as the functional integral
\begin{equation}
Z = \int D[\Phi^{\rm q},\Phi^{\rm c}]\, e^{iS[\Phi^{\rm q},\Phi^{\rm c}]}.
\end{equation}

\begin{figure}
    \centering
    \includegraphics[width=\linewidth]{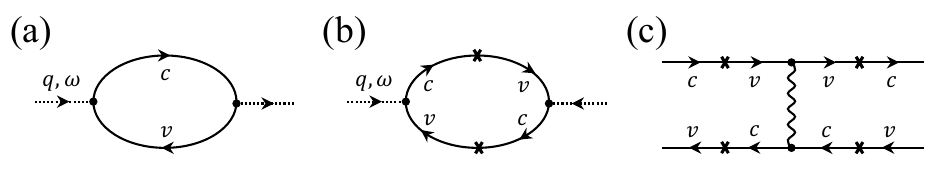}
    \caption{Diagrammatic representations of the effective exciton action (a,b) and an electron-hole scattering process due to interlayer tunneling (c). The solid and dotted curves represent fermion and boson fields respectively, the crosses represent interlayer tunneling, and the wavy line represents Coulomb interaction. (a) and (b) respectively represent the free exciton action $S_0$ and Josephson action $S_J$.} %In the Keldysh formalism, the vertices and fermion lines are $2\times 2$ matrices in Keldysh space; more detailed diagrams are shown in the Supplemental Material.}
    \label{fig:diag}
\end{figure}

Expanding the action in powers of $\Phi$ and in powers of interlayer tunneling we find the leading term
\begin{widetext}
\begin{equation} \label{eq:S0}
S_0[\bar{\Phi},\Phi] = \frac 1A \sum_q \int\frac{d\omega}{2\pi}\, \Tr\bigg[ (\omega-\frac{q^2}{2M}-E_g+E_b+i\gamma) \Phi^{\rm q\dagger}_q(\omega) \Phi^{\rm c}_q(\omega) + {\rm c.c.} + ig(\omega-\mu_x)\coth\frac{\omega-\mu_x}{2T} \Phi^{\rm q\dagger}_q(\omega) \Phi^{\rm q}_q(\omega) \bigg]
\end{equation}
which describes free excitons with energy $E_g-E_b+q^2/2M$ (where the exciton mass $M=m_e^*+m_h^*$) and chemical potential $\mu_x$ at temperature $T$. While the quadratic coefficients take the stated form only in the dilute exciton regime (BEC regime) $|E_g-E_b-\mu_x| \ll E_b$ and in the frequency range $|\omega-\mu_x| \ll E_b$, the overall form of Eq.~\eqref{eq:S0} is general and we expect that our qualitative results apply to a larger parameter regime. %Eq.~\eqref{eq:S0} applies when $\omega \approx \mu_x \approx E_g-E_b$, \textit{i.e.}, for small exciton densities, and for exciton fields with phases rotating in a small range of frequencies determined by the bias voltage. This is the parameter regime that captures the physics of exciton condensation phase transitions. 
The imaginary coefficients $\gamma$ and $g$ describe coupling of excitons to leads. Fluctuation-dissipation theorem implies $\gamma = g(\omega-\mu_x)$. In the absence of interlayer tunneling, the bias voltage $\mu_x$ can be absorbed into $\omega$ and the system is equivalent to an unbiased bilayer with a reduced band gap $E_g-E_b-\mu_x$. Excitons spontaneously form and undergo BEC at low enough temperatures when $E_g-E_b < \mu_x$. Below the transition the exciton fields have semiclassical solutions of the form $\Phi^{\rm c}(t) = |\Phi^{\rm c}| e^{-i\mu_x t}$ with amplitude determined by the ratio of quadratic and quartic coefficients of the action. 

P-wave interlayer tunneling gives rise to a second-order Josephson action of the form
\begin{equation} \label{eq:S_J}
S_J[\bar{\Phi},\Phi] = \frac 1A \sum_{i=1,2} \sum_q \int\frac{d\omega}{2\pi} \left[ -c_J \Phi^{{\rm q},i}_{-q}(-\omega) \Phi^{{\rm c},i}_q(\omega) - c_J \Phi^{{\rm c},i}_{-q}(-\omega) \Phi^{{\rm q},i}_q(\omega) + ig_J \Phi^{{\rm q},i}_{-q}(-\omega) \Phi^{{\rm q},i}_q(\omega) + {\rm c.c.} \right],
\end{equation}
\end{widetext}
in which the intervalley exciton fields $\Phi^1, \Phi^2$ at frequency $\omega$ are coupled to those at frequency $-\omega$. Because of this coupling, the bias voltage $\mu_x$ cannot be absorbed into $\omega$ and the system is out of equilibrium as shown in the Supplemental Material. Diagrammatic representations of $S_0$ and $S_J$ are shown in Fig.~\ref{fig:diag}(a,b). 

If interlayer tunneling is s-wave, a first-order Josephson term proportional to $\Phi^{\rm q}(\omega=0)$ exists. Second-order terms of the form $\Phi(-\omega)\Phi(\omega)$ also exist and the coefficients are equal for all valley components. For p-wave interlayer tunneling \eqref{eq:t_k}, in contrast, angular momentum conservation implies that first-order Josephson terms vanish and that second-order terms are nonzero only for intervalley exciton fields $\Phi^1,\Phi^2$.
The second-order Josephson action \eqref{eq:S_J} produces an energy landscape with explicit dependence on the phase angle $\theta = \arg\Phi$ of the form $E_J\sim \cos 2\theta$ that breaks the $U(1)$ phase symmetry down to $\mathbb{Z}_2$. The interlayer tunneling current satisfies the second-order Josephson relation $I\sim \sin 2\theta$ \cite{sun2021second}.  For an unbiased electron-hole bilayer below the BEC transition, the exciton fields are static and the system picks one of the two preferred phase angles that differ by $\pi$ as the ground state implying Ising-type phase transitions of the exciton fields. Because the Josephson action \eqref{eq:S_J} involves only intervalley exciton fields, intervalley excitons are energetically favored over intravalley excitons by pinning the phase at one of the two preferred phase angles, in agreement with mean-field theory results in the context of InAs/GaSb quantum wells \cite{xue2018time, zeng2022plane}. 

\begin{figure}
\centering
\includegraphics[width=\linewidth]{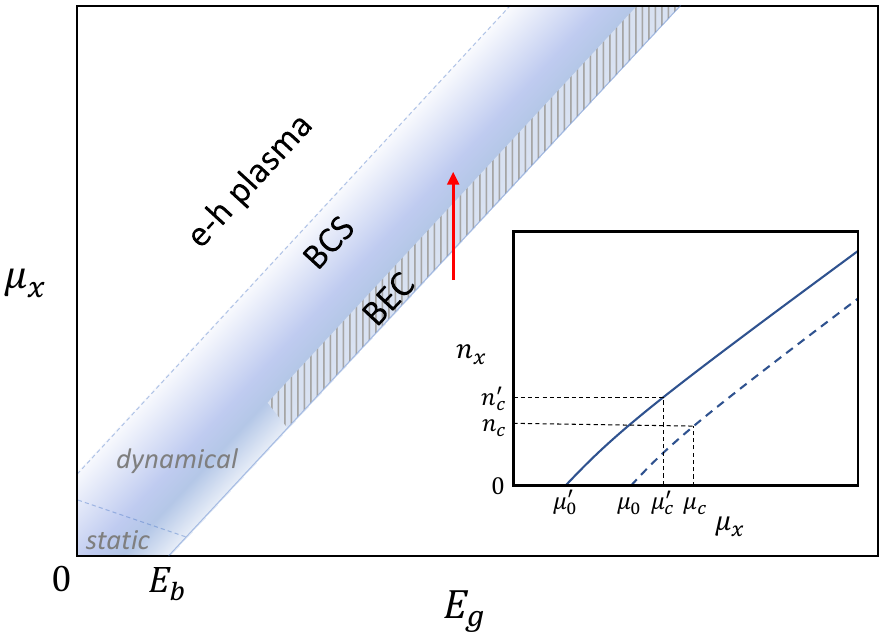}
\caption{Schematic phase diagram of a biased electron-hole bilayer with interlayer tunneling. $E_g$ is the interlayer band gap, $E_b$ is the binding energy of interlayer excitons, and $\mu_x$ is the bias voltage. The system is out of equilibrium when $\mu_x\ne 0$. The blue region represents the region in which exciton condensation occurs, and the color scale represents the strength of excitonic coherence. As the exciton density increases, the condensate undergoes a BEC-BCS crossover \cite{comte1982exciton, nozieres1985bose} and then a Mott transition \cite{mott1949basis, mott1973metal, brinkman1973electron, guerci2019exciton} to an electron-hole plasma. The hatched area represents the large-bias BEC regime in which our theory applies. Inset: schematic plot of exciton density $n_x$ as $\mu_x$ increases along the red arrow. The solid and dashed curves are the $n_x$--$\mu_x$ curves with and without interlayer tunneling respectively. $\mu_0=E_g-E_b$ is the threshold bias voltage for injection of excitons in the absence of interlayer tunneling, $\mu_c$ and $n_c$ are the critical bias voltage and critical density for the occurrence of BKT transition, and $\mu_0'$, $\mu_c'$, and $n_c'$ are the corresponding quantities in the presence of interlayer tunneling.}
\label{fig:phase_diagram}
\end{figure}

{\it Large bias limit.---} In the absence of interlayer tunneling, the phase of the exciton field rotates at a constant frequency $\omega=\mu_x$. Interlayer tunneling leads to a potential landscape that explicitly breaks the $U(1)$ phase symmetry. The interplay between the $U(1)$ symmetry breaking term that traps the phase of the condensate and the bias voltage that drives a rotating phase gives rise to interesting nonequilibrium physics that is different from previous work on driven-dissipative condensates \cite{sieberer2013dynamical, sieberer2016keldysh, dunnett2016keldysh, maghrebi2016nonequilibrium, dalla2013keldysh}. For small bias voltage $\mu_x$, the exciton condensate is a static one with its phase trapped at one of the potential minima. Above a threshold bias voltage $\mu_x\sim c_J$ the condensate becomes a dynamical one with rotating phase. The transition from static to dynamical condensates is schematically shown in Fig.~\ref{fig:phase_diagram}. If the bias voltage is much larger than the Josephson energy scale $c_J$, the phase-dependent energy landscape is swept rapidly by the rotating fields at approximately constant frequency $\omega\approx\mu_x$. Instead of Josephson effects, the Josephson action produces an average effect on the exciton fields and $U(1)$ symmetry is effectively restored.
%In TMDs, the band gap $E_g$ is typically an order of magnitude larger than the binding energy $E_b$, so a large bias voltage $\mu_x > E_g-E_b \gg E_b$ is required to inject electrons and holes into the system. The parameter regime relevant for exciton condensation phase transitions is $\omega \approx \mu_x \approx E_g-E_b$. In other words, the exciton fields of interest have a fast rotating phase. %\CommentVC{Name of quickly rotating phase in Josephson community + Refs?}

To make the above statement more precise, we note that weak interlayer tunneling ($c_J\ll \mu_x$) acts as a small perturbation that does not significantly affect the frequency of phase rotation. Thus the physically active fields are $\Phi(\omega\approx\mu_x)$ with a frequency range determined by $c_J$. The Josephson action $S_J$ couples the physically active fields $\Phi(\omega\approx\mu_x)$ to the frozen degrees of freedom $\Phi(-\omega\approx -\mu_x)$. Since the $\Phi(-\omega)$ fields are trivially gapped, we can integrate them out at the quadratic level and obtain an effective action for the $\Phi(\omega)$ fields:
\begin{widetext}
\begin{equation} \label{eq:S1}
S_1[\bar{\Phi},\Phi] = \frac 1A \sum_{i=1,2} \sum_q \int\frac{d\omega}{2\pi} \left[ \varepsilon \bar{\Phi}^{{\rm q},i}_q(\omega) \Phi^{{\rm c},i}_q(\omega) + {\rm c.c.} + i\lambda \bar{\Phi}^{{\rm q},i}_q(\omega) \Phi^{{\rm q},i}_q(\omega) \right],
\end{equation}
\end{widetext}
where the $\omega$-integral is defined over the small frequency range $|\omega-\mu_x| \lesssim c_J$. Eq.~\eqref{eq:S1} suggests that interlayer tunneling produces an extra contribution to both the c-q and q-q quadratic terms for intervalley exciton fields. The c-q coefficient $\varepsilon>0$ is an effective decrease of the band gap (or enhancement of the exciton binding energy), while the q-q coefficient $\lambda$ implies an effective increase of temperature $\delta T = \lambda/2g$. An order-of-magnitude estimate of the coefficients yields $\varepsilon\sim (mv_t^2)^2 E_b^3/\mu_x E_g^3$ and $\delta T \sim (mv_t^2)^2 E_b^5/\mu_x E_g^5$. 

Physically the action \eqref{eq:S1} originates from the electron-hole scattering process illustrated by the diagram in Fig.~\ref{fig:diag}(c), where an electron and a hole tunnel to the other layer, scatter by interlayer Coulomb potential, and then tunnel back to their original layers. Such scattering process enhances the effective electron-hole interactions and increases the exciton binding energy. For s-wave excitons with p-wave interlayer tunneling, the net contribution is nonzero only when the electron and hole are from opposite valleys so that angular momentum is conserved in the scattering process. Another equivalent point of view \cite{sun2023dynamical} is that p-wave interlayer tunneling leads to a Pondermotive force that favors intervalley excitons in the large-bias and low-density limit. This process breaks the degeneracy between intravalley and intervalley excitons and lowers the degeneracy of the ground state manifold either from $S^1\times S^3$ to $S^1\times S^1$ or from $S^1\times S^2\times S^2$ to $S^1\times \mathbb{Z}_2$, depending on the sign of the exchange quartic term~\cite{wu2015theory} (see Supplemental Material). Because of the repulsion between intravalley and intervalley excitons, the ground state consists of only intervalley excitons even when the bias voltage is above the threshold value for intravalley excitons.

The effective temperature increase that shows up as an extra contribution to the q-q coefficient is physically a fluctuating force on the intervalley exciton fields and breaks the fluctuation-dissipation theorem. In our case it is the $\Phi(-\omega)$ fields that act as an extra fluctuating force on the $\Phi(\omega)$ fields, with coupling strength proportional to interlayer tunneling amplitude. The effective temperature $T_{\rm eff} = T + \delta T$ is the temperature that controls the thermal distribution of intervalley excitons, and is the one that relates response to correlation functions of intervalley exciton fields. The emergence of an effective temperature is common in the Keldysh field theory analysis of driven-dissipative systems \cite{maghrebi2016nonequilibrium, dalla2013keldysh, dunnett2016keldysh, mitra2006nonequilibrium, mitra2005semiclassical}.

{\it Nonequilibrium effects.---} The binding energy of interlayer excitons in few-layer hBN separated TMD bilayers is typically $E_b\sim\SI{100}{meV}$ and decreases with the interlayer distance $d$. The band gap $E_g\sim\SI{1}{eV}$ is an order of magnitude larger than $E_b$, but can be tuned by a displacement field produced by the difference between top and bottom gate voltages. Altogether, the ratio $\delta T/\varepsilon \sim (E_b/E_g)^2 \sim 0.01$ is a small number, which seems to suggest that the increase of effective temperature is a negligible effect.

% \begin{figure}
% \centering
% \includegraphics[width=0.95\linewidth]{n_mu.pdf}
% \caption{Exciton density $n_x$ as a function of bias voltage (exciton chemical potential) $\mu_x$. The solid and dashed curves are the $n_x$--$\mu_x$ curves with and without interlayer tunneling respectively and are shifted horizontally by $\varepsilon$. $\mu_0=E_g-E_b$ is the threshold bias voltage for injection of excitons in the absence of interlayer tunneling, $\mu_c$ and $n_c$ are the critical bias voltage and critical density for the occurrence of BKT transition, while $\mu_0' = \mu_0 - \delta\mu_0$, $\mu_c' = \mu_c - \delta\mu_c$, and $n_c' = n_c + \delta n_c$ (see Eq.~\ref{eq:dmu_c}) are the corresponding quantities in the presence of p-wave interlayer tunneling.}
% \label{fig:n_mu}
% \end{figure}

A finer look at the nonequilibrium effects unveils that, despite the small $E_b/E_g$ ratio, the nonequilibrium dissipative term $\delta T$ can be as important as the $\varepsilon$ term. To see this, we sketch in the inset of Fig.~\ref{fig:phase_diagram} the density of intervalley excitons $n_x$ as a function of the bias voltage $\mu_x$. The gap reduction for intervalley exciton discussed above shifts the $n_x$--$\mu_x$ curve to the left by $\delta \mu_0 = \varepsilon$. Exciton condensation occurs when the temperature is below the Berezinskii-Kosterlitz-Thouless (BKT) transition temperature~\cite{kosterlitz1973ordering,filinov2010berezinskii,fogler2014high,Note1}
\footnotetext[1]{Although Eq.~\eqref{eq_BKT} was obtained for equilibrium systems, since $U(1)$ symmetry is effectively restored in the large bias limit, Eq.~\eqref{eq_BKT} is applicable but with $T$ and $n_x$ modified by nonequilibrium effects (Eq.~\ref{eq:S1}).}
\begin{equation} \label{eq_BKT}
T_{\rm BKT} %= \frac{\pi n_s}{2m_x} 
\approx 1.3\,\frac{n_x}{M},
\end{equation}
%where $n_s$ and $n_x$ are the superfluid density and total density of excitons respectively, and $m_x=2m$ is the exciton mass. 
with $M$ the exciton mass and $n_x$ the exciton density. In other words, the critical exciton density for the occurrence of BKT transition at temperature $T$ is $n_c \approx MT/1.3$. Due to the effective temperature increase, the critical density increases by $\delta n_c \approx M \delta T/1.3$. Since the $n_x$--$\mu_x$ curve is approximately linear at small exciton densities, with the slope approximately given by \cite{wu2015theory,zeng2020electrically} the geometric capacitance $C = e^2 \partial n_x/\partial \mu_x \approx \epsilon/d$ of the bilayer, the critical bias voltage in the presence of interlayer tunneling decreases by
\begin{equation} \label{eq:dmu_c}
%\begin{split}
\delta \mu_c = \delta \mu_0 - \frac{e^2}{C } \delta n_c \approx \varepsilon - \frac{M e^2 d}{1.3\,\epsilon} \delta T,  %\\ &\approx \varepsilon - \frac{m_x e^2 d}{1.3\,\epsilon} \delta T.
%\end{split}
\end{equation}
For a TMD bilayer with a few-layer hBN dielectric spacer, $M\approx m_e$, $d\approx\SI{2}{nm}$, $\epsilon\approx 5\epsilon_0$ (here $m_e$ is the free electron mass and $\epsilon_0$ is the vaccum permittivity), we estimate the prefactor of $\delta T$ in Eq.~\eqref{eq:dmu_c} to be around 73. Since $\varepsilon$ and $\delta T$ also differ by two orders of magnitude, the expression in Eq.~\eqref{eq:dmu_c} can be either positive or negative in realistic systems, and its sign can be tuned by a displacement field that changes the ratio $E_b / E_g$.

%\CommentVC{The above analysis neglects the interactions between intravalley and intervalley excitons. In reality, however, two species of excitons strongly interact by dipole repulsion and exchange interactions. As shown in the Supplemental Material, the ground state consists of only intervalley excitons when interlayer tunneling makes them energetically favorable. Therefore, only intervalley excitons are present in the system even above $\mu_{02}$, and BKT transition of intervalley excitons occurs at the critical bias voltage $\mu_{c1}$.}

{\it Discussion.---} We have shown in this Letter that when interlayer tunneling takes the p-wave form \eqref{eq:t_k}, the degeneracy between intravalley and intervalley excitons is lifted. If a large bias voltage is applied between the electron and hole layers, the $U(1)$ symmetry breaking caused by interlayer tunneling is averaged out by the fast rotating exciton fields. The main effects of interlayer tunneling are the reduction of effective band gap and increase of effective temperature for intervalley excitons. %Surprisingly, these non-equilibrium effects either allow for a predominantly exciton condensate with a predominant inter- or intra-valley character. 

The assumption of p-wave interlayer tunneling \eqref{eq:t_k} is crucial for our results and deserves further discussion. Our theory applies to InAs/GaSb quantum wells and angle-aligned TMD homobilayers with four of the six high-symmetry stackings ($R_h^h, R_h^X, H_h^h$, and $H_h^X$ \cite{rivera2018interlayer,liu2015electronic}), interlayer tunneling is p-wave and our theory is directly applicable.
When interlayer tunneling is s-wave (e.g., TMD homobilayers with $R_h^M$ or $H_h^M$ stacking), a nonzero first-order Josephson term ($\propto \Phi(\omega=0)$) exists for intravalley excitons, leading to nonzero static exciton density even before the condensation transition occurs. While the tunneling-induced static excitons are not coupled to the high-frequency exciton fields at quadratic level, electrostatic repulsion between excitons leads to an effective gap increase for excitons in both valleys.

For TMD heterobilayers or homobilayers with a nonzero twist angle, the two layers form a moir\'e pattern with spatially varying local stacking registry. A proper treatment of general TMD bilayers needs to take account of the momentum shift between conduction and valence bands \cite{bistritzer2011moire,wang2017interlayer,wu2019topological,yu2017moire} and is left for future work. Excitonic coherence between shifted bands leads to density wave states that break translational symmetry \cite{zeng2022plane, rickhaus2021correlated, kogar2017signatures}. In a simple intuitive picture, excitons in a moir\'e potential are localized near one of the high-symmetry stacking sites \cite{wu2018theory, yu2017moire}, and the effects of interlayer tunneling are determined by the local stacking registry.

%As a first approximation, the net effects of spatially varying interlayer tunneling are expected to be a weighted average of the effects of s-wave and p-wave tunneling in different spatial regions. Since s-wave tunneling does not distinguish intravalley and intervalley excitons, the net effects, contributed by p-wave tunneling, are the decrease of effective band gap and increase of effective temperature for intervalley excitons.

{\it Acknowledgements.---}
The authors thank Zhiyuan Sun and Allan MacDonald for helpful discussions. YZ and AJM acknowledge support from Programmable Quantum Materials, an Energy Frontiers Research Center funded by the U.S. Department of Energy (DOE), Office of Science, Basic Energy Sciences (BES), under award DE-SC0019443.   The Flatiron Institute is a division of the Simons Foundation.

\bibliography{BibExciton}

% ------------------------------------------------------------------------ %
% ------------------------------------------------------------------------ %
% ------------------------------------------------------------------------ %

\onecolumngrid
\newpage
\makeatletter 

\begin{center}
\textbf{\large Supplemental material for `` \@title ''} \\[10pt]
% authors \\
% \textit{Institution}
\end{center}
\vspace{20pt}

\setcounter{figure}{0}
\setcounter{section}{0}
\setcounter{equation}{0}

\renewcommand{\thefigure}{S\@arabic\c@figure}
\makeatother

%\twocolumngrid
%\appendix
%\newpage

% ------------------------------------------------------------------------ %
% ------------------------------------------------------------------------ %
% ------------------------------------------------------------------------ %

\section{Derivation of the Keldsyh action for excitons}

\subsection{Keldysh formalism}

\begin{figure}
    \centering
    \includegraphics[width=\linewidth]{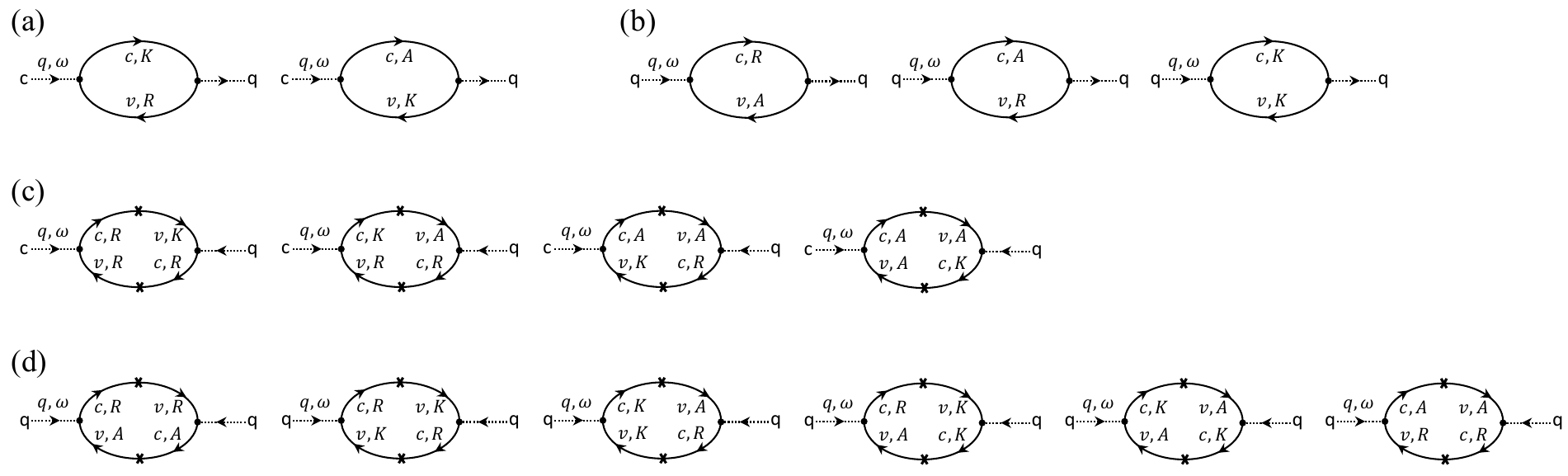}
    %\caption{Diagrammatic representations of the coefficients of (a) $\bar{\Delta}^{\rm c} \Delta^{\rm q}$; (b) $\bar{\Delta}^{\rm q} \Delta^{\rm q}$; (c) $\Delta^{\rm c} \Delta^{\rm q}$; (d) $\Delta^{\rm q} \Delta^{\rm q}$; (e) $\Delta^{\rm q}$. Here the dotted lines represent the bosonic fields $\bar{\Delta},\Delta$ (or the $\bar{\Phi},\Phi$ fields after projecting onto the 1s-exciton basis), solid lines represent Green's functions for electrons (Eqs.~\ref{eq:Green_function}), the wavy lines represent the electromagnetic field, and the crosses represent interlayer tunneling.}
    \caption{Diagrammatic representations of the coefficients of (a) $\bar{\Delta}^{\rm c} \Delta^{\rm q}$; (b) $\bar{\Delta}^{\rm q} \Delta^{\rm q}$; (c) $\Delta^{\rm c} \Delta^{\rm q}$; (d) $\Delta^{\rm q} \Delta^{\rm q}$. Here the dotted lines represent the bosonic fields $\bar{\Delta},\Delta$ (or the $\bar{\Phi},\Phi$ fields after projecting onto the 1s-exciton basis), solid lines represent Green functions for electrons (Eqs.~\ref{eq:Green_function}), and the crosses represent interlayer tunneling.}
    \label{fig:diagrams}
\end{figure}

The central quantity in the Keldysh formalism is the partition function
\begin{equation}
Z = \Tr\left\{ \rho_0 \mathcal{T}_C \exp[-i\int_C dt H(t)] \right\},
\end{equation}
where $C$ is a closed time contour that starts from the distant past $t=-\infty$, proceeds to the distant future $t=\infty$, and then returns to the distant past. $\int_C = \int_{-\infty}^{\infty} + \int_{\infty}^{-\infty}$ is the integral along the closed time contour $C$, and $\mathcal{T}_C$ represents contour ordering along $C$. $\rho_0$ is the density matrix in the infinite past. The partition function is equivalently expressed in the path-integral formalism as
\begin{equation}
Z = \int D[\bar{\psi}^+,\bar{\psi}^-,\psi^+,\psi^-] e^{iS[\bar{\psi}^+,\bar{\psi}^-,\psi^+,\psi^-]} = \int D[\bar{\psi}^1,\bar{\psi}^2,\psi^1,\psi^2] e^{iS_K[\bar{\psi}^1,\bar{\psi}^2,\psi^1,\psi^2]},
\end{equation}
where $\bar{\psi}^{\pm},\psi^{\pm}$ are the fermion fields on the forward ($+$) and backward ($-$) time paths, and in the last expression we have performed a Keldysh rotation
\begin{equation}
\begin{pmatrix} \psi^1 \\ \psi^2
\end{pmatrix} = \frac{1}{\sqrt{2}} \begin{pmatrix} \psi^+ + \psi^- \\ \psi^+ - \psi^-
\end{pmatrix}, \quad
\begin{pmatrix} \bar{\psi}^1 \\ \bar{\psi}^2
\end{pmatrix} = \frac{1}{\sqrt{2}} \begin{pmatrix} \bar{\psi}^+ - \bar{\psi}^- \\ \bar{\psi}^+ + \bar{\psi}^-
\end{pmatrix}
\end{equation}
to eliminate the redundancy in contour space. We choose $\rho_0$ to be the density matrix that describes the equilibrium state of the system without interactions or tunneling at temperature $T$. The Keldysh action then consists of the following parts:
\begin{equation}
S_K = S_c + S_v + S_t + S_{\rm inter} + S_{\rm intra}.
\end{equation}
Here $S_{c/v}$ describes the conduction/valence band electrons coupled to separate leads at temperature $T$ and electrochemical potential $\mu_{c/v}$. In frequency space it has an elegant expression (band index $b=c,v$)
\begin{equation}
S_b = \sum_{\tau k} \int\frac{d\nu}{2\pi} \bar{\psi}_{\tau bk}(\nu) G_{bk}^{-1}(\nu) \psi_{\tau bk}(\nu),
\end{equation}
where $\bar{\psi} = (\bar{\psi}^2, \bar{\psi}^2)$, $\psi = (\psi^1,\psi^2)^T$, and $G$ is a $2\times 2$ matrix of Green functions
\begin{equation}
G_{bk}(\nu) = \begin{pmatrix}
G_{bk}^R(\nu) & G_{bk}^K(\nu) \\
0 & G_{bk}^A(\nu)
\end{pmatrix}.
\end{equation}
The retarded, advanced, and Keldysh Green functions are defined as
\begin{subequations} \label{eq:Green_function}
\begin{align}
& G_{bk}^R(\nu) = 1/(\nu - \xi_{bk} + i\Gamma_b), \\
& G_{bk}^A(\nu) = [G_{bk}^R(\nu)]^*, \\
& G_{bk}^K(\nu) = [G_{bk}^R(\nu) - G_{bk}^A(\nu)]\tanh\frac{\nu-\mu_b}{T}, \label{eq:G_K}
\end{align}
\end{subequations}
where $\Gamma_{c/v}$ is the tunneling rate of the conduction/valence band electrons to the lead, which we take as a constant independent of momentum and frequency. Interlayer tunneling is described by the action
\begin{equation}
S_t= -\sum_{\tau k} \int_C dt_C \left[ t_{\tau k} \bar{\psi}_{\tau ck}(t_C) \psi_{\tau vk}(t_C) + {\rm c.c.} \right] = -\sum_{\tau k} \int\frac{d\nu}{2\pi} \left[ t_{\tau k} \bar{\psi}_{\tau c k}(\nu) \alpha_0 \psi_{\tau vk}(\nu) + {\rm c.c.} \right],
\end{equation}
where the contour time $t_C$ consists of a discrete contour label $C=\pm$ and a continuous time variable $t\in(-\infty,\infty)$, and $\alpha_0$ is the $2\times 2$ identity matrix in Keldysh space. 
% $S_{\rm em}$ describes coupling to an electromagnetic field (neglecting the $A^2$ terms):
% \begin{equation}
% \begin{split}
% S_{\rm em} = -\frac 1A \sum_{\tau kq} \int\frac{d\omega}{2\pi} \int\frac{d\nu}{2\pi} &\bigg\{ \frac km \cdot A_q(\omega) \big[\bar{\psi}_{\tau c,k+q/2}(\nu+\frac{\omega}{2}) \psi_{\tau c,k-q/2}(\nu-\frac{\omega}{2}) - \bar{\psi}_{\tau v,k+q/2}(\nu+\frac{\omega}{2}) \psi_{\tau v,k-q/2}(\nu-\frac{\omega}{2}) \big] \\
% &+ v_t (\tau A_x + iA_y)_{q,\omega} \bar{\psi}_{\tau c,k+q/2}(\nu+\frac{\omega}{2}) \psi_{\tau v,k-q/2}(\nu-\frac{\omega}{2}) + {\rm c.c.} \bigg\}.
% \end{split}
% \end{equation}
$S_{\rm inter}$ and $S_{\rm intra}$ are the interlayer and intralayer parts of Coulomb interaction:
\begin{align}
S_{\rm inter} &= -\frac{1}{A} \sum_{\tau\tau'} \sum_{kk'q} \int_C dt_C V(q) \bar{\psi}_{\tau c,k+q}(t_C) \bar{\psi}_{\tau' v,k'-q}(t_C) \psi_{\tau' vk'}(t_C) \psi_{\tau ck}(t_C), \\
S_{\rm intra} &= -\frac{1}{2A} \sum_{b\tau\tau'} \sum_{kk'q} \int_C dt_C U(q) \bar{\psi}_{\tau b,k+q}(t_C) \bar{\psi}_{\tau' b,k'-q}(t_C) \psi_{\tau' bk'}(t_C) \psi_{\tau bk}(t_C).
\end{align}
By a Hubbard-Stratonovich transformation, $S_{\rm inter}[\bar{\psi},\psi]$ is equivalent to the action (upon integration)
\begin{equation}
S_{\rm inter}[\bar{\Delta},\Delta,\bar{\psi},\psi] = -\frac 1A \sum_{\tau\tau'} \int_C dt_C \left\{\sum_{kk'q} V_{kk'}^{-1} \bar{\Delta}_{kq}^{\tau\tau'}(t_C) \Delta_{k'q}^{\tau\tau'}(t_C) + \sum_{kq} \big[\Delta_{kq}(t_C) \bar{\psi}_{\tau c,k+\frac{m_e^*}{M}q}(t_C) \psi_{\tau' v,k'-\frac{m_h^*}{M}q}(t_C) + {\rm c.c.}\big] \right\},
\end{equation}
where $M=m_e^*+m_h^*$ and $V_{kk'}^{-1}$ is the $(k,k')$-component of the inverse of the interlayer Coulomb interaction matrix $V(k-k')$. After the Keldysh rotation (`c' and `q' stand for `classical' and `quantum' respectively)
\begin{equation}
\begin{pmatrix} \Delta^{\rm c} \\ \Delta^{\rm q}
\end{pmatrix} = \frac{1}{\sqrt{2}} \begin{pmatrix} \Delta^+ + \Delta^- \\ \Delta^+ - \Delta^-
\end{pmatrix}, \quad
\begin{pmatrix} \bar{\Delta}^{\rm c} \\ \bar{\Delta}^{\rm q}
\end{pmatrix} = \frac{1}{\sqrt{2}} \begin{pmatrix} \bar{\Delta}^+ + \bar{\Delta}^- \\ \bar{\Delta}^+ - \bar{\Delta}^-
\end{pmatrix}
\end{equation}
and Fourier transform, $S_{\rm inter}$ becomes
\begin{equation} \label{eq:S_inter_Delta}
\begin{split}
S_{\rm inter} &= -\frac 1A \sum_{\tau\tau'} \sum_{kk'q} \int\frac{d\omega}{2\pi}\,  V_{kk'}^{-1} \bar{\Delta}_{kq}^{\tau\tau'}(\omega) \alpha_1 \Delta_{k'q}^{\tau\tau'}(\omega) \\
&- \frac{1}{\sqrt{2}A} \sum_{\tau\tau'} \sum_{kq} \int\frac{d\omega}{2\pi} \int\frac{d\nu}{2\pi} \left\{ \bar{\psi}_{\tau c,k+\frac{m_e^*}{M}q}(\nu+\frac{\omega}{2}) \big[\Delta_{kq}^{{\rm c},\tau\tau'}(\omega) \alpha_0 + \Delta_{kq}^{{\rm q},\tau\tau'}(\omega) \alpha_1 \big] \psi_{\tau v,k-\frac{m_h^*}{M}q}(\nu-\frac{\omega}{2}) + {\rm c.c.} \right\},
\end{split}
\end{equation}
where $\alpha_1$ is the Pauli-$x$ matrix in Keldysh space. Integrating out the fermion fields, we get an effective action in terms of the pairing fields $\Delta$. Fig.~\ref{fig:diagrams} show the diagrams for the low-order expansions in powers of tunneling and the pairing fields.

\subsection{Classical-quantum quadratic Green function}

The two diagrams in Fig.~\ref{fig:diagrams}(a) represent the following contribution to the action:
\begin{equation}
\frac{i}{2A^2} \sum_{\tau\tau'} \sum_{kq} \int\frac{d\omega}{2\pi} \bar{\Delta}_{kq}^{{\rm c},\tau\tau'}(\omega) \Delta_{kq}^{{\rm q},\tau\tau'}(\omega) \int\frac{d\nu}{2\pi} \left[ G_{c,k+\frac{m_e^*}{M}q}^K(\nu+\frac{\omega}{2}) G_{v,k-\frac{m_h^*}{M}q}^R(\nu-\frac{\omega}{2}) + G_{c,k+\frac{m_e^*}{M}q}^A(\nu+\frac{\omega}{2}) G_{v,k-\frac{m_h^*}{M}q}^K(\nu-\frac{\omega}{2}) \right].
\end{equation}
The $\nu$-integral has an analytic expression when $T\to 0$, $\Gamma_{c,v}\to 0^+$. Assuming $|\mu_{c,v}|<E_g/2$ as is the case for dilute exciton systems, together with the first line of Eq.~\eqref{eq:S_inter_Delta}, the classical-quantum quadratic terms at zeroth order in tunneling sum up to
\begin{equation}
S_{2,0}^{\rm cq}[\bar{\Delta},\Delta] = -\frac{1}{A^2} \sum_{\tau\tau'} \sum_{kq} \int\frac{d\omega}{2\pi}\, \bar{\Delta}_{kq}^{{\rm q},\tau\tau'}(\omega) \left[ \frac{1}{\omega - (\xi_{c,k+\frac{m_e^*}{M}q} - \xi_{v,k-\frac{m_h^*}{M}q})} \Delta_{kq}^{{\rm c},\tau\tau'}(\omega) + A\sum_{k'} V_{kk'}^{-1} \Delta_{k'q}^{{\rm c},\tau\tau'}(\omega) \right] + {\rm c.c.}
\end{equation}
Projecting onto the 1s-exciton basis by the ansatz $\Delta_{kq} = \sum_{k'} V_{kk'} \varphi_{k'} \Phi_q /A$ in which $\varphi_{k'}$ is the wave function for the relative electron-hole motion in a 1s-exciton that satisfies the eigen-mode equation
\begin{equation}
\frac{k^2}{2m}\varphi_k - \frac 1A \sum_{k'} V_{kk'}\varphi_{k'} = -E_b \varphi_k,
\end{equation}
where $m=m_e^* m_h^*/M$ is the reduced mass and $E_b$ is the binding energy of the 1s exciton mode, $S_{2,0}^{\rm cq}$ is equivalently expressed in terms of the $\bar{\Phi},\Phi$ fields as
\begin{equation}
S_{2,0}^{\rm cq}[\bar{\Phi},\Phi] = \frac 1A \sum_{\tau\tau' q} \int\frac{d\omega}{2\pi} (\omega - \frac{q^2}{2M} - E_g + E_b + i\gamma) \bar{\Phi}_q^{{\rm q},\tau\tau'}(\omega) \Phi_q^{{\rm c},\tau\tau'}(\omega) + {\rm c.c.},
\end{equation}
in the limit $\omega\approx E_g-E_b+q^2/2M$. An imaginary coefficient $i\gamma$ that describes coupling to the reservoirs is included for completeness. Lowest-order expansion in $\Gamma_{c,v}$ gives $\gamma\propto \Gamma_c \Gamma_v (\omega-\mu_c+\mu_v)$.

\subsection{Quantum-quantum quadratic Keldysh term}

The diagrams in Fig.~\ref{fig:diagrams}(b) represent the integrals
\begin{equation} \label{eq:int_GG_qq}
\begin{split}
&\frac i2 \sum_{\rm qq} \int GG = \frac i2 \int\frac{d\nu}{2\pi} (G_{c+}^R G_{v-}^A + G_{c+}^A G_{v-}^R + G_{c+}^K G_{v-}^K) \\
&= -\frac i2 \coth\frac{\omega-\mu_c+\mu_v}{2T} \int\frac{d\nu}{2\pi} (G_{c+}^R-G_{c+}^A) (G_{v-}^R-G_{v-}^A) \big(\tanh\frac{\nu+\frac{\omega}{2}-\mu_c}{2T} - \tanh\frac{\nu-\frac{\omega}{2}-\mu_v}{2T}\big),
\end{split}
\end{equation}
where the $\pm$ subscripts represent the momentum-frequency arguments $(k\pm q/2,\nu\pm\omega/2)$. In the last expression we have made use of Eq.~\eqref{eq:G_K} and hyperbolic trigonometric identities. In the limit $\omega\approx \mu_c-\mu_v$, the Green's functions are approximately constants in the narrow range $\nu\in (\mu_c-\omega/2, \mu_v+\omega/2)$ and one obtains
\begin{equation}
\frac i2 \sum_{\rm qq} \int GG = \frac{2i\Gamma_c \Gamma_v/\pi}{(\xi_{c+} -\mu_c)^2 (\xi_{v-} -\mu_v)^2} (\omega-\mu_x) \coth\frac{\omega-\mu_x}{2T},
\end{equation}
where $\xi_{b,\pm} = \xi_{b,k\pm q/2}$ and $\mu_x=\mu_c-\mu_v$ is the exciton chemical potential. The quantum-quantum quadratic action then reads
\begin{equation}
S_{2,0}^{\rm qq}[\bar{\Phi},\Phi] = \frac 1A \sum_{\tau\tau' q} \int\frac{d\omega}{2\pi}\, ig(\omega-\mu_x)\coth\frac{\omega-\mu_x}{2T}\, \bar{\Phi}_q^{{\rm q},\tau\tau'}(\omega) \Phi_q^{{\rm q},\tau\tau'}(\omega).
\end{equation}
To zeroth order in $q$, the coefficient $g$ is a constant with an order-of-magnitude estimate $g\sim \Gamma_c \Gamma_v/E_b^2$.

$S_{2,0} = S_{2,0}^{\rm cq} + S_{2,0}^{\rm qq}$ describes a system of free excitons without interlayer tunneling. Effective thermodynamic equilibrium requires the relation $\gamma = g(\omega-\mu_x)$.

\subsection{Classical-quantum Josephson term}

$\Delta\Delta$ and $\bar{\Delta}\bar{\Delta}$ terms arise when the $U(1)$ phase symmetry of the pairing fields is explicitly broken by the tunneling term. Fig.~\ref{fig:diagrams}(c) shows the diagrams for the $\Delta^{\rm q}\Delta^{\rm c}$ terms at quadratic order in tunneling, which consist of the integrals
\begin{equation}
\frac i2 \sum_{\rm cq} \int GtGGtG = \frac i2 t_{\tau+}^* t_{\tau'-}^* \int\frac{d\nu}{2\pi} (G_{c+}^R G_{v+}^K G_{c-}^R G_{v-}^R + G_{c+}^K G_{v+}^A G_{c-}^R G_{v-}^R + G_{c+}^A G_{v+}^A G_{c-}^R G_{v-}^K + G_{c+}^A G_{v+}^A G_{c-}^K G_{v-}^A),
\end{equation}
where $t_{\tau\pm} = t_{\tau,k\pm q/2}$. At zero temperature and $\Gamma_{c,v}\to 0^+$, $|\mu_{c,v}|<E_g/2$, one obtains
\begin{equation}
\frac i2 \sum_{\rm cq} \int GtGGtG = -\frac{1}{(\xi_{c+}-\xi_{v+}) (\xi_{c-}-\xi_{v-})} \left( \frac{1}{\omega-\xi_{c+}+\xi_{v-}} - \frac{1}{\omega+\xi_{c-}-\xi_{v+}} \right) t_{\tau+}^* t_{\tau'-}^*.
\end{equation}
In terms of the $\Phi$ fields, only the intervalley $\tau'=\bar{\tau}\equiv -\tau$ terms survive in the $k$-integral and give rise to the Josephson term
\begin{equation}
S_{2,2}^{\rm cq}[\bar{\Phi},\Phi] = -\frac 1A \sum_{\tau q} \int\frac{d\omega}{2\pi}\, c_J \Phi_{-q}^{{\rm q},\tau\bar{\tau}}(-\omega) \Phi_q^{{\rm c},\bar{\tau}\tau}(\omega) + {\rm c.c.}
\end{equation}
As an order-of-magnitude estimate, $c_J\sim mv_t^2 E_b^2/E_g^2$ when $\omega\approx E_g-E_b$.

\subsection{Quantum-quantum Josephson term}

The diagrams in Fig.~\ref{fig:diagrams}(d) represent the integrals
\begin{equation}
\begin{split}
\frac i2 \sum_{\rm qq} \int GtGGtG = \frac i2 t_{\tau+}^* t_{\tau'-}^* \int\frac{d\nu}{2\pi} &(G_{c+}^R G_{v+}^R G_{c-}^A G_{v-}^A + G_{c+}^R G_{v+}^K G_{c-}^R G_{v-}^K + G_{c+}^K G_{v+}^A G_{c-}^R G_{v-}^K \\
&+ G_{c+}^R G_{v+}^K G_{c-}^K G_{v-}^A + G_{c+}^K G_{v+}^A G_{c-}^K G_{v-}^A + G_{c+}^A G_{v+}^A G_{c-}^R G_{v-}^R).
\end{split}
\end{equation}
The Green's function integrals can be rewritten in a more illuminating form analogous to Eq.~\eqref{eq:int_GG_qq}:
\begin{equation}
\begin{split}
\sum_{\rm qq} \int GGGG = &-\coth\frac{\omega-\mu_x}{2T} \int\frac{d\nu}{2\pi} (G_{c+}^R-G_{c+}^A) G_{v+}^A G_{c-}^R (G_{v-}^R-G_{v-}^A) \left( \tanh\frac{\nu+\frac{\omega}{2}-\mu_c}{2T} - \tanh\frac{\nu-\frac{\omega}{2}-\mu_v}{2T} \right) \\
&-\coth\frac{\omega}{2T} \int\frac{d\nu}{2\pi} (G_{c+}^R-G_{c+}^A) G_{v+}^A (G_{c-}^R-G_{c-}^A) G_{v-}^A \left( \tanh\frac{\nu+\frac{\omega}{2}-\mu_c}{2T} - \tanh\frac{\nu-\frac{\omega}{2}-\mu_c}{2T} \right) \\
&-\coth\frac{\omega}{2T} \int\frac{d\nu}{2\pi} G_{c+}^R (G_{v+}^R-G_{v+}^A) G_{c-}^R (G_{v-}^R-G_{v-}^A) \left( \tanh\frac{\nu+\frac{\omega}{2}-\mu_v}{2T} - \tanh\frac{\nu-\frac{\omega}{2}-\mu_v}{2T} \right) \\
&-\coth\frac{\omega+\mu_x}{2T} \int\frac{d\nu}{2\pi} G_{c+}^R (G_{v+}^R-G_{v+}^A) (G_{c-}^R-G_{c-}^A) G_{v-}^A \left( \tanh\frac{\nu+\frac{\omega}{2}-\mu_v}{2T} - \tanh\frac{\nu-\frac{\omega}{2}-\mu_c}{2T} \right).
\end{split}
\end{equation}
The corresponding action in terms of $\Phi$ fields is
\begin{equation}
S_{2,2}^{\rm qq}[\bar{\Phi},\Phi] = \frac 1A \sum_{\tau q} \int\frac{d\omega}{2\pi}\, ig_J \Phi_{-q}^{{\rm q},\tau\bar{\tau}}(-\omega) \Phi_q^{{\rm q},\bar{\tau}\tau}(\omega) + {\rm c.c.}
\end{equation}
In the physically interesting case $\omega\approx\mu_x\gtrsim E_b$, the coefficient is approximately a constant $g_J\sim (\Gamma_c+\Gamma_v)mv_t^2 E_b^2/E_g^3$.

% \subsection{Coupling to electromagnetic field}

% The diagrams in Fig.~\ref{fig:diagrams}(e) sum up to
% \begin{equation}
% \frac{v_t (\tau A_x - iA_y)_{-q,-\omega}}{\omega - \xi_+ - \xi_-} - \frac{v_t (\tau k_x - ik_y)}{\omega - \xi_+ - \xi_-} \left(\frac{1}{\xi_+} + \frac{1}{\xi_-} \right) \frac{k}{2m}\cdot A_{-q}(-\omega).
% \end{equation}
% Neglecting $q$-dependence of coefficients, the above expression can be rewritten as
% \begin{equation}
% A_{-q}(-\omega)\cdot \frac{\partial}{\partial k} \frac{v_t (\tau k_x - ik_y)}{\omega - 2\xi_k} - \omega\, \frac{v_t (\tau k_x - ik_y)}{\xi_k (\omega - 2\xi_k)^2}\, \frac{k}{m} \cdot A_{-q}(-\omega).
% \end{equation}
% The first term is an inconsequential momentum shift. Integrating over $k$, the second term leads to the action
% \begin{equation}
% S_{1,1}^{\rm q}[\bar{\Phi},\Phi] = -\frac 1A \sum_{\tau q} \int\frac{d\omega}{2\pi}\, c_{\rm em}\, \omega (\tau A_x - iA_y)_{-q,-\omega} \Phi^{{\rm q},\tau\tau}_q(\omega) + {\rm c.c.}
% \end{equation}
% with coefficient $c_{\rm em}\sim v_t \sqrt{mE_b}/E_g$.

%\subsection{Quartic terms}

\section{Effective action for nonequilibrium excitons}

\subsection{Large bias limit}

In the following we express the $2\times 2$ matrix $(\Phi^{\tau\tau'})$ in terms of the 4-component spinor $(\Phi^{\mu})$ defined by $\Phi^{\tau\tau'} = (\sum_{\mu} \Phi^{\mu}\tau_{\mu}/\sqrt{2})^{\tau\tau'}$. The physically interesting parameter regime is $\omega\approx\mu_x \approx E_g-E_b$, but the Josephson terms couples the intervalley excitons at $\omega\approx\mu_x$ with those at $\omega\approx -\mu_x$. Assuming that $\mu_x$ is much larger than the cutoff scale of $\omega$, we separate the quadratic action into three parts corresponding to the parameter ranges $\omega\approx\pm\mu_x$ and the Josephson coupling terms. At $\omega\approx\mu_x$,
\begin{equation}
S_0[\bar{\Phi},\Phi] = \frac 1A \sum_q \int\frac{d\omega}{2\pi}\, \Tr\bigg[ (\omega-\frac{q^2}{2M}-E_g+E_b+i\gamma) \Phi^{\rm q\dagger}_q(\omega) \Phi^{\rm c}_q(\omega) + {\rm c.c.} + ig(\omega-\mu_x)\coth\frac{\omega-\mu_x}{2T} \Phi^{\rm q\dagger}_q(\omega) \Phi^{\rm q}_q(\omega) \bigg].
\end{equation}
At frequency $-\omega\approx -\mu_x$,
\begin{equation}
S_0'[\bar{\Phi},\Phi] = \frac 1A \sum_q \int\frac{d\omega}{2\pi}\, \Tr\left[ -c_2' \Phi^{\rm q\dagger}_{-q}(-\omega) \Phi^{\rm c}_{-q}(-\omega) + {\rm c.c.} + ig' \Phi^{\rm q\dagger}_{-q}(-\omega) \Phi^{\rm q}_{-q}(-\omega) \right],
\end{equation}
where the coefficients are approximately constants $c_2'\sim E_b \mu_x/E_g$, $g'\sim \Gamma_c \Gamma_v E_b \mu_x/E_g^3$. The Josephson terms couple the $\Phi(\omega\approx\mu_x)$ and $\Phi(-\omega\approx -\mu_x)$ fields:
\begin{equation}
S_J[\bar{\Phi},\Phi] = \frac 1A \sum_{i=1,2} \sum_q \int\frac{d\omega}{2\pi} \left[ -c_J \Phi^{{\rm q},i}_{-q}(-\omega) \Phi^{{\rm c},i}_q(\omega) - c_J \Phi^{{\rm c},i}_{-q}(-\omega) \Phi^{{\rm q},i}_q(\omega) + ig_J \Phi^{{\rm q},i}_{-q}(-\omega) \Phi^{{\rm q},i}_q(\omega) + {\rm c.c.} \right].
\end{equation}
The above three integrals are all defined in a narrow range ($\lesssim c_J$) near $\omega\approx\mu_x$. Integrating out the $\Phi(-\omega\approx -\mu_x)$ fields, we get an effective action for $\Phi(\omega\approx\mu_x)$ fields. The intravalley pairing fields $\Phi^{0,3}$ remain unaffected due to the absence of coupling terms, while the intervalley pairing fields receive an extra contribution
\begin{equation}
S_1[\bar{\Phi},\Phi] = \frac 1A \sum_{i=1,2} \sum_q \int\frac{d\omega}{2\pi} \left[ \varepsilon \bar{\Phi}^{{\rm q},i}_q(\omega) \Phi^{{\rm c},i}_q(\omega) + {\rm c.c.} + i\lambda \bar{\Phi}^{{\rm q},i}_q(\omega) \Phi^{{\rm q},i}_q(\omega) \right],
\end{equation}
where the coefficients $\varepsilon=c_J^2/c_2'\sim (mv_t^2)^2 E_b^3/E_g^4$ and $\lambda = g'c_J^2/c_2'^2 \sim \Gamma_c \Gamma_v (mv_t^2)^2 E_b^3/E_g^6$. Physically the coefficient $\varepsilon$ is an effective decrease of the energy gap for intervalley excitons, while $\lambda$ implies an effective increase of temperature
\begin{equation}
\delta T = \lambda/2g \sim (mv_t^2)^2 E_b^5/E_g^6.
\end{equation}

\subsection{Violation of fluctuation-dissipation theorem (FDT)}

In this section we assume that $\mu_x\approx E_g-E_b$ is small and demonstrate how FDT is violated when $\mu_x\ne 0$. The quadratic action for intervalley excitons can be expressed in the matrix form
\begin{equation} \label{eq:S_matrix}
S[\bar{\Phi},\Phi] = \frac 1A \sum_q \int_0^{\Lambda} \frac{d\omega}{2\pi} \begin{pmatrix}
\bar{\Phi}_q^{\rm c} & \bar{\Phi}_q^{\rm q} & \Phi_{-q}^{\rm c} & \Phi_{-q}^{\rm q}
\end{pmatrix}
\begin{pmatrix}
0 & \omega-E_q-i\gamma_+ & 0 & -c_J \\
\omega-E_q+i\gamma_+ & i\gamma_+ \coth\frac{\omega-\mu_x}{2T} & -c_J & ig_J \\
0 & -c_J & 0 & -\omega-E_q+i\gamma_- \\
-c_J & -ig_J & -\omega-E_q-i\gamma_- & i\gamma_- \coth\frac{\omega+\mu_x}{2T}
\end{pmatrix}
\begin{pmatrix}
\Phi_q^{\rm c} \\ \Phi_q^{\rm q} \\ \bar{\Phi}_{-q}^{\rm c} \\ \bar{\Phi}_{-q}^{\rm q}
\end{pmatrix},
\end{equation}
where $\Lambda$ is a frequency cutoff, $E_q = E_g-E_b+q^2/4m$ is the exciton energy, and $\gamma_{\pm} = g(\omega\mp\mu_x)$. The subscripts $q$ of the fields in the basis vectors are shorthand for the momentum and frequency labels $(q,\omega)$. The values of $c_J$ and $g_J$ are not the same as those calculated above for the large bias limit and can have nontrivial dependence on $q$ and $\omega$, but their precise values are not important for the following analysis.

The quadratic action \eqref{eq:S_matrix} can in principle be block-diagonalized into two $2\times 2$ blocks by a Bogoliubov transformation. The q-q component of each block would then be a mixture of $\coth\frac{\omega\pm\mu_x}{2T}$ terms (unless $c_J=0$), and the sum takes the standard coth form only when $\mu_x=0$. Here we use a simpler method to demonstrate the violation of FDT. As in the last section, we integrate out the $\Phi(-\omega)$ fields and obtain an effective action for the $\Phi(\omega>0)$ fields. The result is
\begin{equation} \label{eq:S_matrix}
S'[\bar{\Phi},\Phi] = \frac 1A \sum_q \int_0^{\Lambda} \frac{d\omega}{2\pi} \begin{pmatrix}
\bar{\Phi}_q^{\rm c} & \bar{\Phi}_q^{\rm q}
\end{pmatrix}
\begin{pmatrix}
0 & \omega-E_q-i\gamma_+ + \frac{c_J^2}{\omega+E_q+i\gamma_-} \\
\omega-E_q+i\gamma_+ + \frac{c_J^2}{\omega+E_q-i\gamma_-} & i\gamma_+ \coth\frac{\omega-\mu_x}{2T} + \frac{ic_J^2 \gamma_-}{(\omega+E_q)^2+\gamma_-^2} \coth\frac{\omega+\mu_x}{2T}
\end{pmatrix}
\begin{pmatrix}
\Phi_q^{\rm c} \\ \Phi_q^{\rm q}
\end{pmatrix}.
\end{equation}
It is easy to check that FDT is satisfied when either $\mu_x$ or $c_J$ vanishes but violated when both are nonzero.

\section{Ground state manifold}

To investigate the ground state manifold of the exciton system, in this section we work with the real-time Schrodinger field theory, which is equivalent to the linear-in-$\Phi^{\rm q}$ part of the Keldysh nonequilibrium field theory but does not contain information about thermal occupation. Up to quartic order in $\Phi$, the system without interlayer tunneling is described by the Lagrangian
\begin{equation}
L_0 = \Tr[\Phi^{\dagger}(i\partial_t - E_0) \Phi] - \frac{c_H}{2} [\Tr(\Phi^{\dagger}\Phi)]^2 - c_X \Tr[(\Phi^{\dagger}\Phi)^2].
\end{equation}
Here the $q$ labels in $\Phi$ are omitted, with the assumption that only the $q=0$ fields are condensed, and $E_0 = E_{q=0} = E_g-E_b$. The first quartic term, with coefficient $c_H>0$, is the Hartree term that comes from dipole repulsion of interlayer excitons. The other quartic term describes exchange interactions and can be either positive or negative. Mean-field calculations suggest \cite{wu2015theory} that $c_X>0$ for small interlayer distance and $c_X<0$ for large interlayer distance. In the latter case $c_H$ is always large enough to ensure stability. Defining exciton density $\rho = \Tr(\Phi^{\dagger}\Phi)$, the exchange quartic term can be expressed as
\begin{equation}
\Tr[(\Phi^{\dagger}\Phi)^2] = \frac 12 [\rho^2 + (\Phi_0^* \bm{\Phi} + {\rm c.c.} + i\bm{\Phi}^* \times \bm{\Phi})^2],
\end{equation}
where we defined the complex 3-vector $\bm{\Phi} = (\Phi_1,\Phi_2,\Phi_3)$. Assuming that $\Phi(t)$ takes the form $\Phi(t) = e^{-i\mu t}\Phi$ with time-independent $\Phi$, the Lagrangian is then
\begin{equation}
L_0 = (\mu-E_0)\rho - \frac{c_H+c_X}{2} \rho^2 - \frac{c_X}{2} (\Phi_0^* \bm{\Phi} + {\rm c.c.} + i\bm{\Phi}^* \times \bm{\Phi})^2.
\end{equation}
To find the saddle point of $L_0$, we make use of the overall phase degree of freedom to parameterize $\Phi_0=r$ by a real number and $\bm{\Phi} = \bm{u}+i\bm{v}$ by two real 3-vectors. The goal is now to minimize or maximize the length of the vector
\begin{equation}
\bm R = (\Phi_0^* \bm{\Phi} + {\rm c.c.} + i\bm{\Phi}^* \times \bm{\Phi})/2 = r\bm u - \bm{u}\times\bm{v},
\end{equation}
at fixed $\rho$, depending on the sign of $c_X$, and then choose the optimal $\rho$ to maximize $L_0$. Interlayer tunneling introduces an extra term
\begin{equation}
L_1 = \varepsilon (u_1^2+u_2^2+v_1^2+v_2^2)
\end{equation}
with a small positive constant $\varepsilon$. Below we study the ground state manifold with and without $L_1$, in the case of positive and negative $c_X$.

\subsection{$c_X>0$}

When $c_X>0$, $|\bm R|$ needs to be minimized in the ground state, and this is achieved when $\bm u = 0$. The ground state $\Phi = e^{i\theta} (r,i\bm v)$ is parameterized by a real number $r$ and a real 3-vector $\bm v$ that form a 3-sphere $r^2+\bm{v}^2=\rho$, and an overall phase $\theta$. Note that the point $(\theta,r,\bm v)$ is identified with the point $(\theta+\pi,-r,-\bm v)$, so the ground state manifold is $S^1\times S^3$ but the range of $\theta$ is $[0,\pi)$.

The extra term $L_1$ favors intervalley pairing states with $r=v_3=0$, so in this case $v_1$ and $v_2$ form a circle $v_1^2+v_2^2=\rho$ and the ground state manifold is a torus $T^2 = S^1\times S^1$.

\subsection{$c_X<0$}

When $c_X<0$, we need to maximize the quantity
\begin{equation}
|\bm R|^2 = r^2 \bm u^2 + |\bm u\times\bm v|^2 = \bm u^2 (r^2 + \bm v^2) = \bm u^2 (\rho - \bm u^2),
\end{equation}
where the second equality assumes $\bm u\perp \bm v$ as required for maximization. The last expression is maximized when $\bm u^2 = \rho/2$ or equivalently $r^2 + \bm v^2 = \rho/2$. The ground state $\Phi = e^{i\theta} (r,\bm u + i\bm v)$ is parameterized by an overall phase $\theta$, a real 3-vector $\bm u$ on a 2-sphere with radius $\sqrt{\rho/2}$, and another real 3-vector on the same sphere with $r$ its component parallel to $\bm u$ and $\bm v$ the perpendicular component. The ground state manifold is locally $S^1\times S^2\times S^2$, but as before, the point $(\theta,r,\bm u,\bm v)$ is identified with the point $(\theta+\pi,-r,-\bm u,-\bm v)$.

To maximize $L_0+L_1$, $r=0$ and $\bm u,\bm v$ are a set of orthogonal vectors in 1-2 plane with equal fixed length $\sqrt{\rho/2}$. The overall phase angle $\theta$ acts as an in-plane rotation of $\bm u$ and $\bm v$. The ground state can be written as $\Phi = \sqrt{\rho/2} e^{i\theta} (0,1,\pm i,0)$. The ground state manifold is thus $S^1 \times \mathbb{Z}_2$, parameterized by a phase angle $\theta$ and a sign factor.

%\CommentVC{In this last part, you have assumed the overall $U(1)$ phase to implement rotation in the $xy$ plane, while fir the other, this effect was double counted. }

\end{document}